\documentclass[a4paper,twosides]{article}
\usepackage{fontspec}
\usepackage{makecell}

\usepackage{multicol,multirow,graphicx,fancyhdr,epstopdf}
\usepackage{amsmath,amsfonts,amssymb,bm,upgreek,mathrsfs,ccmap,mathcomp}
\usepackage[pagewise,switch,columnwise]{lineno}
\usepackage[compress,nospace]{cite}
\usepackage[dvipsnames]{xcolor}
\usepackage{CPL-2023}
\usepackage{longtable}
\usepackage{capt-of}

\newcommand{\cplyear}{2026} \newcommand{\cplvol}{xx}
 \newcommand{\cplpagenumber}{xxxxxx}
 \newcommand{\cplpage}{\cplpagenumber-\thepage}

\begin{document}

\twocolumn[

\vspace*{-4mm}
\begin{center}

\par\vspace{3mm}
\large\bf{\boldmath{Targeted search for eccentric supermassive binary black holes in OJ 287 and nearby galaxy clusters with PPTA DR3}}

\par\vspace{3mm}

\normalsize \rm{}
Shi-Yi Zhao$^{1,2}$, %
Xingjiang Zhu$^{1*}$, %
Jacob Cardinal Tremblay$^{3,4}$, %
Yiqin Chen$^{1,2}$, %
Ma\l{}gorzata Cury\l{}o$^{5,6}$, %
Shi Dai$^{7}$,
Valentina Di Marco$^{8}$, %
Pratyasha Gitika$^{6,9}$,
George Hobbs$^{7}$, %
Simon C.-C. Ho$^{6,10,11}$,
Xiao-Song Hu$^{1,2,12}$, %
Agastya Kapur$^{7,13}$, %
Wenhua Ling$^{7}$, %
Richard N. Manchester$^{7}$,
Saurav Mishra$^{6,7,11}$, %
Daniel J.~Reardon$^{6,11}$, %
Christopher J~Russell$^{14}$, %
Ryan M.~Shannon$^{11}$, %
Sharon Mary Tomson$^{3,4}$,
Jingbo Wang$^{15}$, %
Shuangqiang Wang$^{7,16}$, %
Andrew Zic$^{6,7}$

\par\vspace{3mm}

$^{1}$Institute for Frontier in Astronomy and Astrophysics \& Faculty of Arts and Sciences, Beijing Normal University, Zhuhai 519087, China

$^{2}$School of Physics and Astronomy, Beijing Normal University, Beijing 100875, China

$^{3}$Max Planck Institute for Gravitational Physics (Albert Einstein Institute), 30167 Hannover, Germany

$^{4}$Leibniz Universität Hannover, 30167 Hannover, Germany

$^{5}$School of Physics and Astronomy, Monash University, Clayton VIC 3800, Australia

$^{6}$ARC Centre for Excellence for Gravitational Wave Discovery (OzGrav), Hawthorn, VIC 3122, Australia

$^{7}$Australia Telescope National Facility, CSIRO, Space \& Astronomy, PO Box 76, Epping, 1710, NSW, Australia

$^{8}$School of Physics, University of Melbourne, Parkville, VIC 3010, Australia

$^{9}$Center for Gravitation, Cosmology and Astrophysics, Department of Physics, University of Wisconsin–Milwaukee, P.O. Box 413, Milwaukee, WI 53201, USA.

$^{10}$Research School of Astronomy and Astrophysics, The Australian National University, Canberra, ACT 2611, Australia 

$^{11}$Centre for Astrophysics and Supercomputing, Swinburne University of Technology, Hawthorn, VIC 3122, Australia

$^{12}$State Key Laboratory of Radio Astronomy and Technology, Shanghai Astronomical Observatory, CAS, Shanghai 200030, China

$^{13}$School of Mathematical and Physical Sciences, Faculty of Science and Engineering, Macquarie University, Sydney, NSW 2109, Australia

$^{14}$CSIRO Scientific Computing, Australian Technology Park, Locked Bag 9013, Alexandria, NSW 1435, Australia

$^{15}$Institute of Optoelectronic Technology, Lishui University, Lishui, 323000, China

$^{16}$Xinjiang Astronomical Observatory, Chinese Academy of Sciences, Urumqi, Xinjiang 830011, P. R. China

\end{center}
\vskip 1.5mm

\small{\narrower We perform Bayesian targeted searches for continuous gravitational waves from eccentric supermassive binary black holes (SMBBHs) using the Parkes Pulsar Timing Array third data release (PPTA DR3). Six electromagnetically motivated sky directions are analyzed, including the blazar OJ~287 and five nearby galaxy clusters (Virgo, Fornax, Norma, Hercules, and Coma). No significant signals are found. For OJ 287, by explicitly incorporating orbital eccentricity (up to $e_0 = 0.8$) to robustly capture signal power spread across multiple harmonics, we constrain the total binary mass to $M_{\rm tot} < 5.25
\times 10^{10} M_{\odot}$ (95\% credible level).
We also place upper limits on the chirp mass of potential SMBBHs residing in galaxy clusters. By combining these limits with independent black hole mass estimates, we place novel constraints on the allowed binary mass ratios for potential hosts such as M87 and NGC~4889. Specifically, our results exclude binaries with mass ratios $q \gtrsim 10^{-2}$ at around 10 nHz for these massive systems, effectively ruling out equal-mass black hole mergers in the sampled parameter space. These findings demonstrate the growing power of pulsar timing arrays to probe SMBBH populations.
\par}\vskip 3mm

\normalsize\noindent{\narrower{KEY WORDS: Gravitational Waves, Pulsar Timing, Suppermassive Black Holes, Multi-messenger Astronomy}

\par}\vskip 5mm
]

\footnotetext{\hspace*{-5.4mm}$^{*}$Corresponding authors. Email: zhuxj@bnu.edu.cn

\noindent\copyright\,{\cplyear}
\href{http://www.cps-net.org.cn}{Chinese Physical Society} and
\href{http://www.iop.org}{IOP Publishing Ltd}}

\textbf{Introduction.}
The direct detection of gravitational waves (GWs) has inaugurated a new era of multi-messenger astrophysics. In the nanohertz frequency band, pulsar timing arrays (PTAs) probe the dynamical Universe by searching for correlated perturbations in the arrival times of millisecond pulsars\ucite{Detweiler1979,HellingsDowns1983,PPTA13PASA}. Following years of tightening constraints, multiple PTA collaborations have recently reported strong evidence for a GW background characterized by the Hellings-Downs angular correlation\ucite{Agazie2023GWB,Antoniadis2023EPTADR2GWB,Reardon2023PPTAGWB,Xu2023CPTADR1GWB,mpta2025gw}. This landmark discovery, which may be attributed to the collective hum of cosmological populations of SMBBHs, marks the beginning of nanohertz GW astronomy. The next major goal for PTAs is the identification of individual SMBBHs via continuous GWs (CGWs), which would offer unique insights into galaxy mergers and binary evolution.

Complementary to all-sky searches\ucite{NG15cw,Falxa2023IPTADR2CW,Antoniadis2024EPTADR2CW,Zhao2025PPTADR3AllSkyCW} that survey the entire volume for unknown sources, targeted searches informed by electromagnetic (EM) observations focus on promising candidate systems or dense galactic environments. By fixing parameters such as sky location and luminosity distance based on EM data, targeted searches significantly reduce the parameter space, potentially enhancing detection sensitivity for specific sources\ucite{pptadr3_3c66b}. These focused studies allow us to test specific binary hypotheses and place astrophysically meaningful constraints on potential host systems, even in the absence of a detection.

A critical factor in these searches is orbital eccentricity, which plays a significant role in the binary evolution at nanohertz frequencies. Unlike the circularized orbits expected in the late inspiral, binaries in the PTA band can sustain significant eccentricity due to environmental interactions, such as stellar scattering, gas drag, or hierarchical triplet dynamics\ucite{Sesana2013CQG, Blaes2002ApJ}. Since eccentricity spreads signal power across multiple harmonics and alters the waveform phase evolution, neglecting it can lead to signal power being missed and parameter estimation biases\ucite{zhu15MNsingle,Taylor2016ecc,ChenWang2022}. Incorporating eccentric waveforms is therefore essential for a faithful reconstruction of the binary evolution and mitigates the risk of model misspecification.

In this work, we present a targeted search for eccentric SMBBHs using the PPTA DR3 dataset\ucite{Zic2023PPTADR3,Reardon2023PPTANull}. We focus on two classes of high-priority targets selected to maximize detection probability. First, we consider the blazar OJ 287. Its well-known quasi-periodic optical outbursts have made it a leading candidate for a SMBBH, with precessing orbital models providing one successful framework for predicting flare times\ucite{Valtonen2016OJ287,Dey2018OJ287}. Second, we target the central regions of five nearby, massive galaxy clusters: Virgo, Fornax, Norma, Hercules, and Coma. These dense environments host rich populations of massive galaxies and brightest cluster galaxies (BCGs), making them promising sites for SMBBHs whose nanohertz GW emission could fall in the PTA band. Their proximity and high dynamical mass make them ideal laboratories for constraining the local SMBBH population.

To perform this search, we employ \texttt{GWecc} waveform model\ucite{Susobhanan2020,Susobhanan2023} to conduct fully Bayesian analyses that consistently incorporate eccentric inspirals, including both Earth and pulsar terms. We compare our results against standard quasi-circular waveforms to isolate the impact of eccentricity and derive astrophysical upper limits on the binary mass and mass ratio for potential hosts in our targeted regions.

\medskip

\textbf{Data and Methods.}
We utilize the PPTA DR3 dataset\ucite{Zic2023PPTADR3}, which contains high-precision timing observations for millisecond pulsars collected over a baseline of up to 18 years. Following the methodology of Refs. \cite{Reardon2023PPTAGWB} and \cite{Zhao2025PPTADR3AllSkyCW}, we select a high-sensitivity subset of 30 pulsars for this analysis. All timing models are referred to the TT(BIPM2020) timescale and use the DE440 Solar System ephemeris. 
Our noise model accounts for pulsar-specific white noise, spin red noise, dispersion measure (DM) variations, and a common red noise (CRN) process\ucite{Reardon2023PPTANull}. The pulsar-specific time-correlated noise components are modeled as power-law Gaussian processes:
\begin{equation}
P_{\rm red}(f;A,\gamma)=\frac{A^2}{12\pi^2}\left(\frac{f}{f_{\rm yr}}\right)^{-\gamma}\,{\rm yr}^3,
\end{equation}
with priors $\pi(\log_{10}A)=U[-18,-11]$ and $\pi(\gamma)=U[0,7]$, consistent with the PPTA DR3 single-pulsar noise analyses\ucite{Reardon2023PPTANull}. In the present targeted search, these time-correlated noise parameters are sampled jointly with the CGW signal, while the white-noise parameters are fixed to their maximum-likelihood values from the single-pulsar noise analyses.

We perform targeted CGW searches toward six directions informed by EM observations: the blazar OJ 287 and the centers of the Virgo, Fornax, Norma, Hercules, and Coma galaxy clusters. To reduce parameter space, we fix each target’s sky location and luminosity distance to EM-derived values (Table~\ref{tab:targets}).
For OJ 287, we additionally fix the GW frequency based on its proposed 12-year orbital timescale\ucite{Sillanpaa1988,Dey2018OJ287}.

A CGW induces a fractional frequency shift (redshift) $z_p(t)$ for pulsar $p$, which is obtained by projecting the metric perturbation onto the pulsar--Earth line of sight. The corresponding timing residual is
\begin{equation}
R_p(t)=\int_{t_0}^{t} [z_p(t')-z_p(t'-\Delta_p)] \, {\rm d}t',
\label{eq:redshift_residual}
\end{equation}
where $t$ denotes the observation time at the Solar System barycenter, $t_0$ is an arbitrary reference epoch (noting that any constant offset introduced by the choice of $t_0$ is absorbed by the timing-model fit), and $\Delta_p$ is the geometric delay between the Earth term and pulsar term
\begin{equation}
\Delta_p = \frac{D_p}{c}\left(1-\cos\mu\right),
\label{eq:delay}
\end{equation}
where $D_p$ is the pulsar distance and $\mu$ is the angular separation between the GW source and pulsar with respect to the observer.

The GW strain can be written as
\begin{equation}
h(t)=F_+\,h_+(t)+F_\times\,h_\times(t),
\label{eq:antenna}
\end{equation}
where $h$ is the GW strain, $F_{+}$ and $F_{\times}$ are the PTA antenna pattern functions. Defining
\begin{equation}
s_{+,\times}(t)\equiv \int_{t_0}^{t} h_{+,\times}(t')\,{\rm d}t',
\end{equation}
the residual can be written as
\begin{equation}
\begin{aligned}
R_p(t)=&\left(F_+\cos2\psi-F_\times\sin2\psi\right)
\left[s_+(t)-s_+(t_p)\right] \\
&+\left(F_+\sin2\psi+F_\times\cos2\psi\right)
\left[s_\times(t)-s_\times(t_p)\right],
\end{aligned}
\label{eq:residual_explicit}
\end{equation}
where $s_{+,\times}(t)$ and $s_{+,\times}(t_p)$ are the Earth-term and pulsar-term contributions, respectively, $t_{p}=t-\Delta_{p}$, and $\psi$ is the GW polarization angle.

For the GWecc model, the quadrupolar-order timing-response functions are
\begin{equation}
s_+(t)=S(t)\left[(1+\cos^2\iota)\left(\mathcal{P}\cos2\omega-\mathcal{Q}\sin2\omega\right)
+\sin^2\iota\,\mathcal{R}\right],
\end{equation}
\begin{equation}
s_\times(t)=2S(t)\cos\iota\left(\mathcal{P}\sin2\omega+\mathcal{Q}\cos2\omega\right),
\end{equation}
where $S(t)$ is the overall time-dependent PTA signal amplitude, $\iota$ is the orbital inclination, $\omega$ is the argument of periastron, and $\mathcal{P}$, $\mathcal{Q}$, and $\mathcal{R}$ are time-dependent functions of the eccentric anomaly and eccentricity. More detailed expressions for the eccentric CGW signal model and its post-Newtonian orbital evolution can be found in Ref. \cite{Agazie2024GWECC}.

Thus, unlike the quasi-circular case, the eccentric waveform depends explicitly on the evolving orbital configuration and redistributes signal power among multiple harmonics. 
We consider two signal models: (i) eccentric inspiral (\texttt{GWecc}), which computes $h_{+,\times}(t)$ for binaries with initial orbital eccentricity $e_0$; (ii) a quasi-circular model (\texttt{cw\_delay}) included in the \texttt{enterprise\_extensions} software package\ucite{enterprise_extensions}. 

The post-fit timing residuals are modeled as
\begin{equation}
\delta \mathbf{t}
= \mathbf{M}\boldsymbol{\epsilon}
+ \mathbf{n}_{\rm PSRN}
+ \mathbf{n}_{\rm CRN}
+ \mathbf{s},
\label{eq:pta_model}
\end{equation}
where $\mathbf{M}\boldsymbol{\epsilon}$ denotes the (linearized) timing-model contribution, $\mathbf{n}_{\rm PSRN}$ denotes pulsar noise, $\mathbf{n}_{\rm CRN}$ is a CRN process, and $\mathbf{s}$ is the deterministic CGW signal. Given the weak evidence for HD correlations in PPTA DR3 ($\mathcal{B}_{\rm HD}\simeq 1.5$), we consider only an uncorrelated CRN here \ucite{Reardon2023PPTAGWB}, This treatment is a simplifying assumption and should be regarded as a limitation of the present analysis. If a weak HD-correlated background is already present in the data, neglecting spatial correlations could in principle affect the inferred CGW posteriors and upper limits, especially at the lowest frequencies. Quantifying that effect is beyond the scope of the present work. Nevertheless, given the low evidence for HD correlations in PPTA DR3, we do not expect the omission of spatial correlations to qualitatively change the main conclusions of this targeted search. A full treatment including spatially correlated common processes will be important in future analyses with more sensitive PTA datasets.
We evaluate a Gaussian likelihood,
\begin{equation}
\ln \mathcal{L}
=
-\frac{1}{2}\left(\delta\mathbf{t}-\mathbf{s}\right)^{\!\top}
\mathbf{C}^{-1}
\left(\delta\mathbf{t}-\mathbf{s}\right)
-\frac{1}{2}\ln\det(2\pi\mathbf{C}),
\label{eq:gaussian_like}
\end{equation}
where $\mathbf{C}$ is the noise covariance including white noise, red noise, and CRN components.

We sample the posterior using the \texttt{PTMCMCSampler} with parallel tempering\ucite{PTMCMCSampler,Vousden2016PT}.
Priors for the eccentric search include uniform distributions for $\log_{10} S_0$ (with $S_0 \equiv S(t_0)$, the PTA signal amplitude at the reference epoch, in seconds) over $[-11,-6]$, symmetric mass ratio $\eta$ over $[10^{-3},0.25]$.
For the GW frequency, we fix $f_{\rm GW}$ to the EM-informed value for OJ~287, while for the five galaxy clusters we adopt a log-uniform prior $\log_{10} f_{\rm GW}$ in $[-9,-7]$.
We adopt the pulsar distance priors from Ref. \cite{Zhao2025PPTADR3AllSkyCW}. However, to be conservative regarding systematic errors in electron density models, we inflate the uncertainties of all DM-derived distances by a factor of two.
Crucially, for the \texttt{GWecc} model, we impose a physical validity restriction $V(\boldsymbol{\Theta})\in\{0,1\}$. This prior rejects binary configurations that would merge within the 18-year observation span, or for which the quasi-Keplerian waveform construction fails over the data span, ensuring self-consistency in the waveform generation.
The effective prior is thus the product of the sampling priors and this validity function.

We sample the initial eccentricity $e_0$ uniformly over $[10^{-3}, 0.8]$. While higher eccentricities can theoretically exist, the quasi-Keplerian framework and harmonic summations underlying the GWecc model are optimized for $e_0 \le 0.8$, ensuring both waveform accuracy and computational tractability.
Beyond eccentricity, the choice of observable mass parameters differs between our targets. For the blazar OJ 287, we parameterize our search in terms of the total mass $M_{\rm tot}$ and symmetric mass ratio $\eta$. This enables a direct comparison with the well-established EM binary models\ucite{Dey2018OJ287}. For the galaxy cluster targets, we place constraints on the chirp mass $\mathcal{M}$, defined as
\begin{equation}
\mathcal{M}=\frac{(m_1m_2)^{3/5}}{(m_1+m_2)^{1/5}}=M_{\rm tot}\eta^{3/5}\, ,
\end{equation}
where $m_1$ and $m_2$ are component masses of the binary system.

Evidence for a CGW is assessed via the Bayes factor $\mathcal{B}$ between hypotheses $\mathcal{H}_1$ (CRN+CGW) and $\mathcal{H}_0$ (CRN only).
We compute $\mathcal{B}$ using the Savage–Dickey density ratio\ucite{Dickey1971,VerdinelliWasserman1995,Wagenmakers2010SDDR}, which estimates the ratio of posterior to prior density at zero signal amplitude.
In the absence of a detection ($\mathcal{B} \lesssim 20$), we report 95\% credibility upper limits on the chirp mass $\mathcal{M}$ and the intrinsic strain amplitude $h_{0}$, defined here as:
\begin{equation}
h_0 = \frac{2(G\mathcal{M})^{5/3}}{c^4 D_L}\,(\pi f_{\rm GW})^{2/3},
\end{equation}
where $D_L$ is the luminosity distance to the GW source.

\begin{table}[t]
\centering
\caption{Sky locations and luminosity distances for the six targeted sources \ucite{Valtonen2016OJ287, Dey2018OJ287, oj287dist, Mei2007, Blakeslee2009, Mutabazi2014, comaher_dist}; for OJ~287 we additionally fix the GW frequency $f_{0}$. We adopt the \texttt{enterprise}/\texttt{GWecc} sky convention with $\phi=\alpha$ and $\cos\theta=\sin\delta$，where $\alpha$ and $\delta$ are the right ascension and declination, respectively.}
\label{tab:targets}
\small
\setlength{\tabcolsep}{3pt}   
\begin{tabular}{lcccc}
\hline
Target & $\cos\theta$ & $\phi~[\mathrm{rad}]$ & $D_L~[\mathrm{Mpc}]$ & $\log_{10}(f_0/\mathrm{Hz})$ \\
\hline
OJ~287    & 0.3438  & 2.334 & 1600 & $-8.277$ \\
Virgo     & 0.2146  & 3.276 & 16.5 & -- \\
Fornax    & $-0.5800$ & 0.9530 & 20 & -- \\
Norma     & $-0.8735$ & 4.252 & 68 & -- \\
Coma      & 0.4692  & 3.403 & 95 & -- \\
Hercules  & 0.3048  & 4.212 & 151 & -- \\
\hline
\end{tabular}
\end{table}

\textbf{Results.} 
Our targeted Bayesian searches find no statistically significant evidence for CGWs from eccentric SMBBHs in any of the six targeted directions. Below, we detail the results for the individual blazar OJ 287 and the five galaxy clusters, presenting Bayesian model comparisons, posterior-derived constraints, and astrophysical upper limits.

\textit{1. OJ~287}

We performed two separate searches toward OJ 287: one using the eccentric \texttt{GWecc} waveform and another using the quasi-circular \texttt{cw\_delay} waveform as a benchmark. In both cases, the sky position, luminosity distance, and GW frequency were fixed by EM constraints (Table \ref{tab:targets}).

We compute the Bayes factor $\mathcal{B}$ comparing the signal hypothesis ($\mathcal{H}_{\rm CRN+OJ287}$, which includes a CGW signal) and the noise hypothesis ($\mathcal{H}_{\rm CRN}$).
The results are:
\begin{equation}
\mathcal{B} =
\begin{cases}
0.82, & \text{eccentric},\\
0.94, & \text{circular}.
\end{cases}
\end{equation}
Values consistent with unity indicate no evidence for a CGW signal from OJ 287 in the PPTA DR3 data under either waveform model.

\begin{figure}[t]
    \centering
    \includegraphics[width=0.98\linewidth]{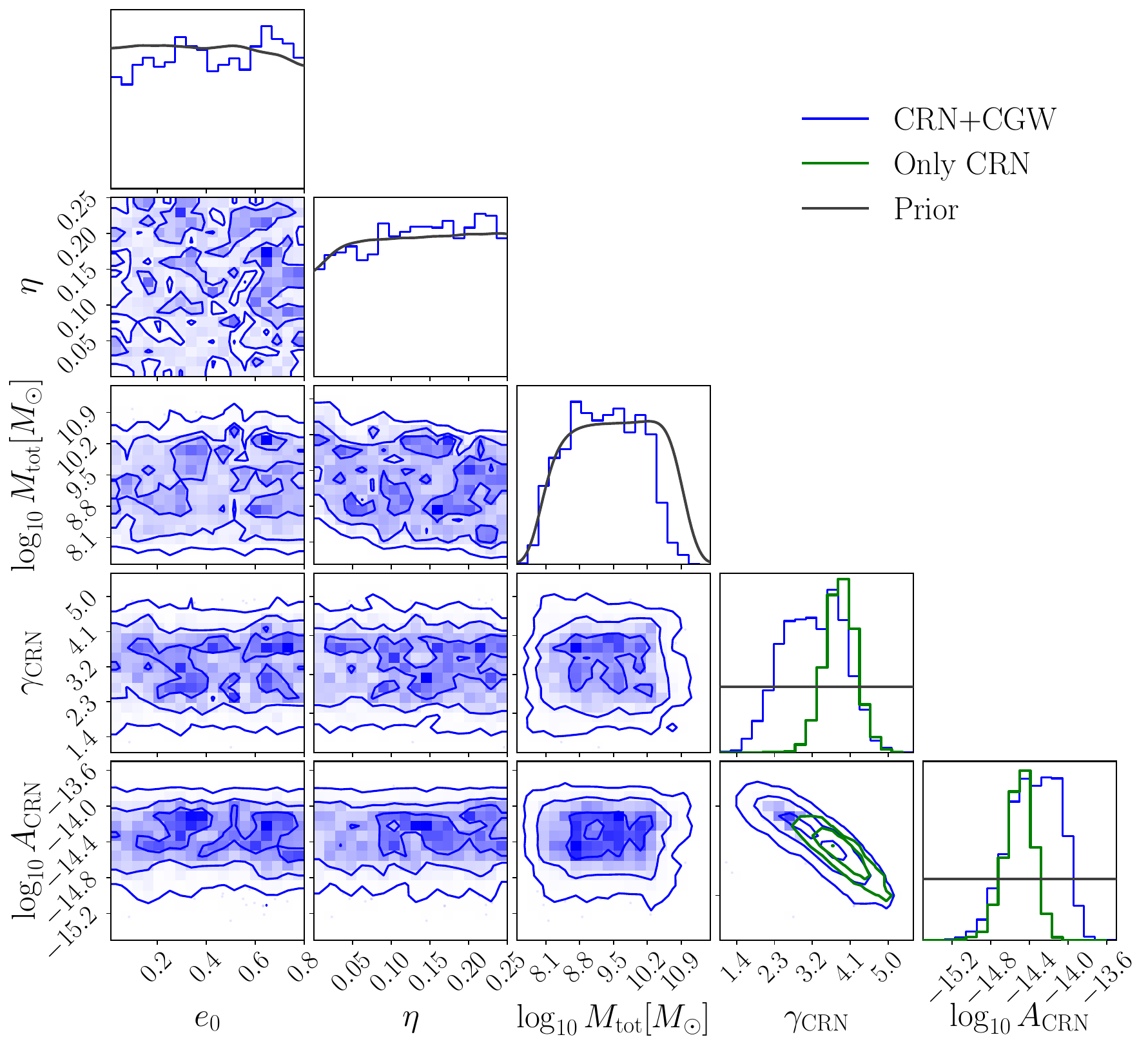}
    \caption{Posterior distributions for the eccentric CGW parameters and CRN parameters from the targeted search toward OJ 287.
    Blue contours and histograms correspond to the signal model $\mathcal{H}_{\rm CRN+OJ287}$, while green shows the noise-only model $\mathcal{H}_{\rm CRN}$. The priors are plotted in gray in each one-dimensional marginalized distribution panel. }
    \label{fig:oj287_corner}
\end{figure}

The joint posterior distributions for the eccentric CGW parameters and the CRN hyperparameters are shown in Figure \ref{fig:oj287_corner}.
We note that the posteriors for all CGW parameters (including total mass $M_{\rm tot}$, eccentricity $e_0$, and mass ratio $\eta$) remain broadly consistent with their validity-restricted priors, with the very high mass region being excluded by the data. This prior-dominated behavior across the parameter space is the expected signature of a non-detection.
Furthermore, the inferred CRN parameters under the signal model ($\mathcal{H}_{\rm CRN+OJ287}$) show good agreement with those from the noise-only model ($\mathcal{H}_{\rm CRN}$), with any differences contained well within the posterior uncertainties. This overall consistency indicates that the introduction of a deterministic CGW component does not significantly alter the explanation of the data provided by the CRN process, visually corroborating the Bayesian model comparison result.

\begin{figure}[t]
    \centering
    \includegraphics[width=0.98\linewidth]{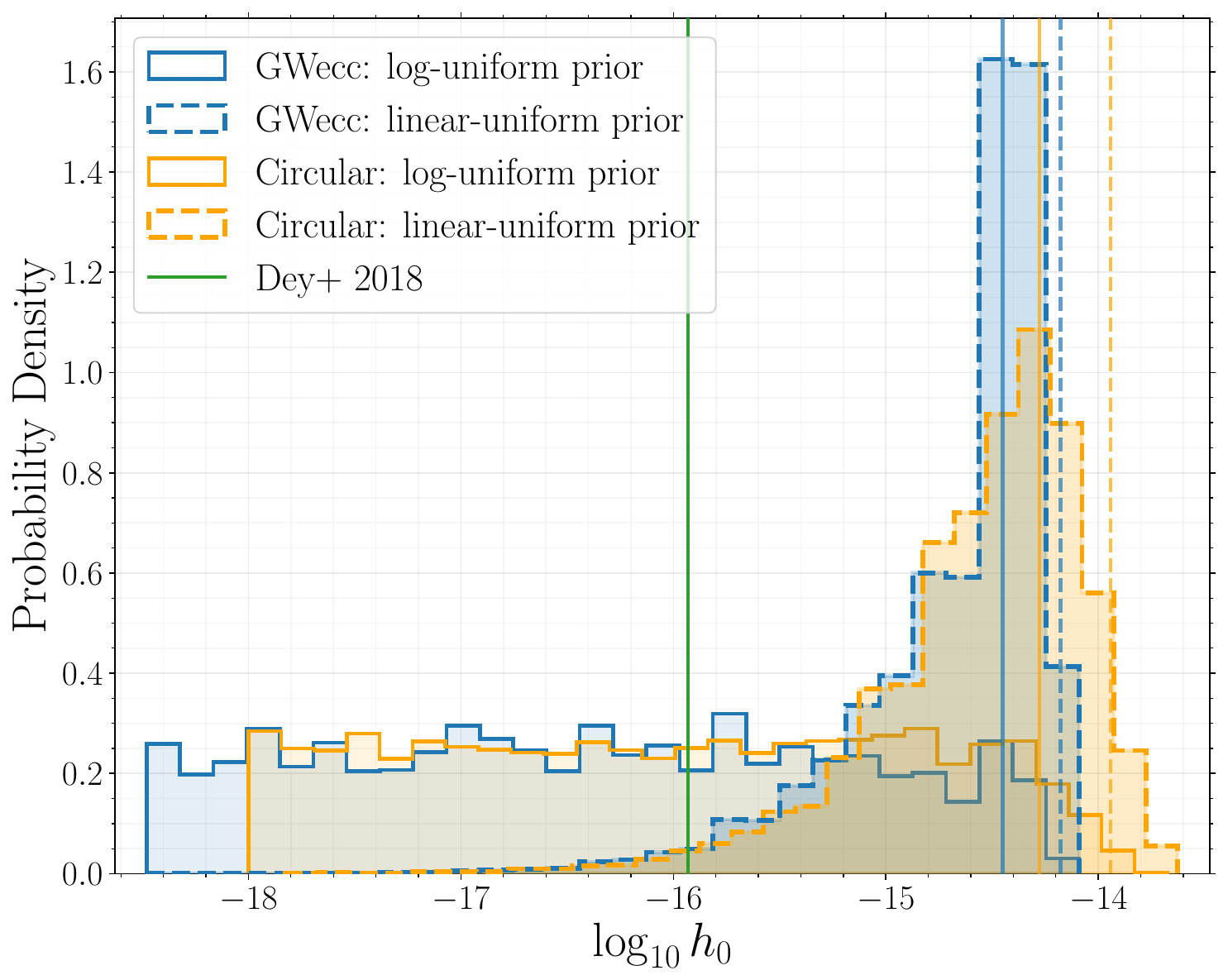}
    \caption{Posterior distributions of the GW characteristic strain amplitude $\log_{10}h_0$. Results are shown for both the eccentric (\texttt{GWecc}) and circular waveform models. For each model, posteriors derived under the searching (log-uniform) prior (solid lines) and the limiting (linear-uniform) prior (dashed lines) are displayed. The corresponding 95\% credibility upper limits are indicated by vertical lines. The predicted strain ($\log_{10}h_0=-15.93$) from the EM binary model of Ref. \cite{Dey2018OJ287} is marked by the green vertical line, lying significantly below the PTA sensitivity. Note that the apparent peaks in the linear-uniform-prior curves are artifacts of the prior reweighting, not evidence of a true signal.}
    \label{fig:oj287_h0}
\end{figure}

Figure~\ref{fig:oj287_h0} presents the posterior distributions for the characteristic strain amplitude $\log_{10}h_0$. The eccentric analysis yields marginally tighter upper limits than the quasi-circular benchmark. The 95\% credible upper limits are $\log_{10}h_0<-14.18$ and $\log_{10}h_0<-13.94$ for the eccentric and circular model, respectively.
For context, the strain amplitude predicted by the binary model of Ref. \cite{Dey2018OJ287} ($\log_{10}h_0=-15.93$) is significantly below our upper bound.


\begin{figure}[t]
    \centering\includegraphics[width=0.98\linewidth]{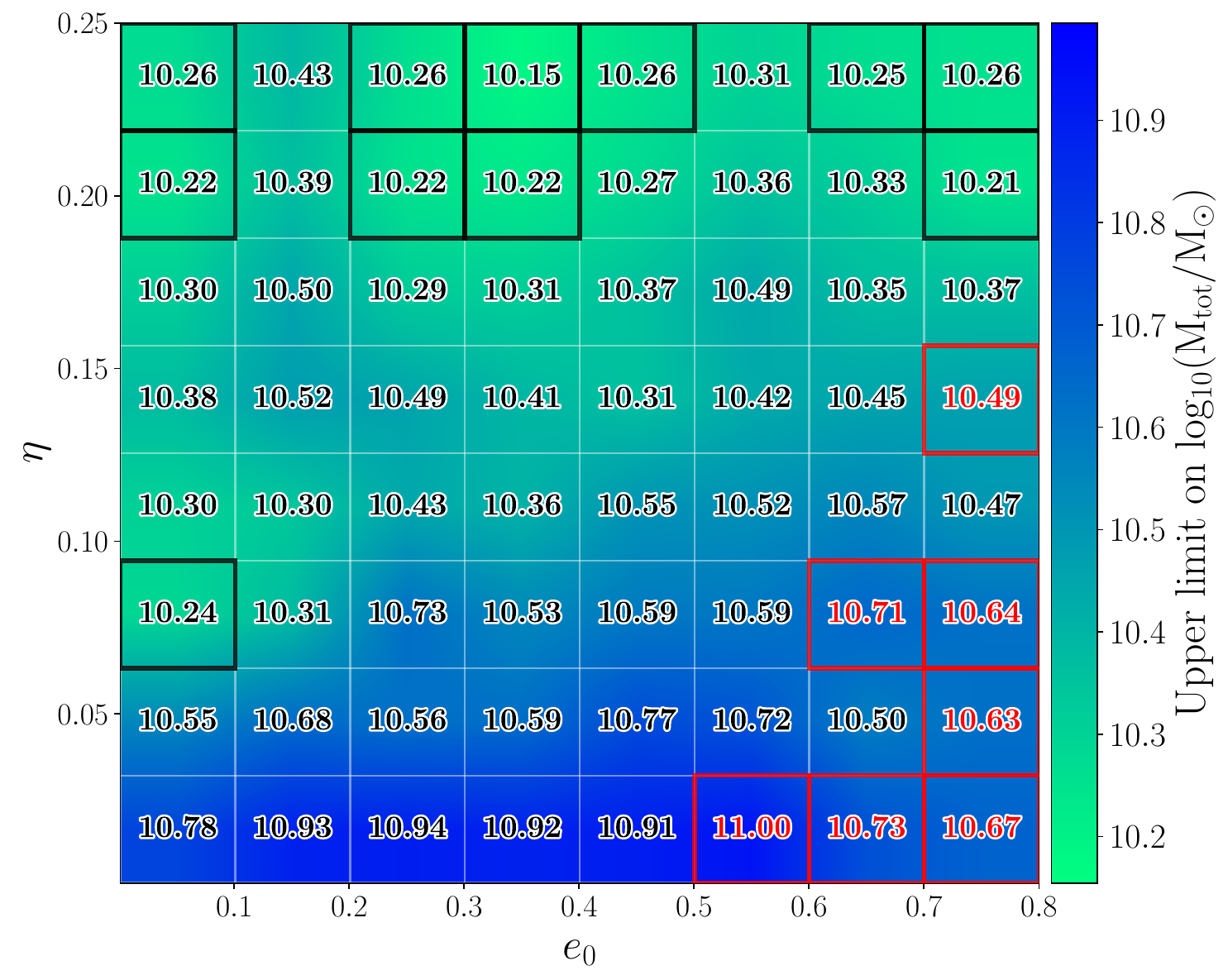}
    \caption{95\% credibility upper limits on the binary total mass across the $(e_0,\eta)$ parameter space for OJ 287. Each grid represents a fixed pair of initial eccentricity $e_0$ and symmetric mass ratio $\eta$, with the number indicating the derived upper limit on $\log_{10}(M_{\rm tot}/M_\odot)$. Red boxes mark pixels where the posterior upper limit differs from the (validity-restricted) prior upper limit by less than 5\%, indicating that the constraint is prior-dominated rather than informed by the data. Black boxes indicate where the PTA upper limit falls below the EM estimate ($\log_{10}M_{\rm tot}=10.27$)\ucite{Dey2018OJ287}, which also infers $e_0=0.657$ and $\eta=0.008$.} 
    \label{fig:oj287_ul_grid}
\end{figure}

A key element of our eccentric search is the use of a validity-restricted prior for the \texttt{GWecc} waveform. This prior ensures the binary orbit evolves in a physically self-consistent manner over the full observational timespan\ucite{Agazie2024GWECC}.
To distinguish between constraints informed by data versus those imposed by this prior, we apply a prior-dominance diagnostic on an $(e_0,\eta)$ grid, as shown in Figure~\ref{fig:oj287_ul_grid}.
This reveals that the prior strongly dominates only at high eccentricities ($e_0\gtrsim 0.6$; red boxes).
This behavior is expected because highly eccentric systems radiate GWs more efficiently\ucite{Blanchet2014LRR,Peters1964}, leading to rapid orbital evolution that is more likely to violate the stability requirement over the 18-year baseline.
It is also worth mentioning that, while the GWecc-based search formally allows initial eccentricities up to $e_0=0.8$, constraints in the high-eccentricity region (particularly for $e_0 \gtrsim 0.6$ in our analysis) should be interpreted with caution. In this regime, binaries become more relativistic and evolve more rapidly, making the results increasingly dependent on the validity conditions imposed to ensure the quasi-Keplerian waveform construction remains self-consistent over the data span. Our reliance on the validity prior in this regime reflects both the rapid physical evolution of the orbit and the inherent applicability limits of the current waveform model.

Across the broad data-informed region of parameter space in Figure~\ref{fig:oj287_ul_grid}, our analysis yields meaningful upper limits.
When compared with the electromagnetically estimated total mass ($\log_{10}M_{\rm tot}=10.27$), our PTA limit falls below this value for low-eccentricity, near-equal-mass configurations, approximately $e_0 \lesssim 0.1$ or $\eta \gtrsim 0.2$ (i.e., $q \gtrsim 0.38$), as indicated by the black boxes in Figure~\ref{fig:oj287_ul_grid}.
This demonstrates how targeted PTA searches can already test specific binary properties in well-motivated systems, even in the absence of a direct detection.

When comparing these values, it is crucial to recognize that our upper limit of $M_{\rm tot} < 5.25 \times 10^{10} M_{\odot}$ is a direct observational constraint derived from the CGW search, whereas the estimate of $1.85 \times 10^{10} \, M_{\odot}$ from Ref. \cite{Dey2018OJ287} is based on a specific binary geometry and the precise modeling of optical flares to infer the system's total mass.
Currently, our PTA observations do not improve upon the EM estimate in the specific parameter space of $e_0 \approx 0.657$ and $\eta \approx 0.008$, as our sensitivity in this high-eccentricity regime is limited by physical validity priors.
However, we expect that with future PPTA and International Pulsar Timing Array (IPTA) data releases, PTA constraints will eventually surpass the estimate of EM-derived models, providing a crucial independent test of the OJ 287 binary hypothesis.

\textit{2. Galaxy clusters}

We perform targeted Bayesian searches toward the central regions of five nearby, massive galaxy clusters: Virgo, Fornax, Norma (Abell 3627), Hercules (Abell 2151), and Coma.
These environments are prime targets for several reasons: they host dense populations of massive galaxies and brightest cluster galaxies (BCGs), are statistically likely to contain SMBBHs emitting nanohertz GWs, and have well-characterized properties from EM observations. For context, Virgo is the nearest massive cluster anchored by the well-studied elliptical M87, a prime candidate with an independently measured supermassive black hole\ucite{safarzadeh2019M87,1994ApJ...435L..35H}. Fornax is a compact, nearby cluster dominated by the massive central galaxy NGC 1399, which is often taken as its BCG\ucite{Wagner1991NGC1399}. Coma is a dense cluster containing the giant elliptical NGC 4889, one of the most massive known black hole hosts\ucite{Sanders2014Coma}. Hercules is dynamically young and irregular\ucite{Huang1996A2151}, suggesting recent galaxy interactions that could have led to SMBBH formation. Finally, Norma is an extremely massive cluster in the Great Attractor region, with a high concentration of galaxies that statistically favors SMBBH formation through mergers\ucite{Woudt2008Norma,Mutabazi2014}.


For all five clusters, the Bayes factors comparing the signal hypothesis (CRN+CGW) to the noise-only hypothesis (CRN) are consistent with or below unity (Table~\ref{tab:bf_clusters}), indicating no statistically significant evidence for an individual CGW source in any targeted direction.
The sole instance of a mild preference ($\mathcal{B}=1.36$ for Coma, using the eccentric model) is far below the threshold for a detection.
We therefore treat all cluster results as non-detections and proceed to derive astrophysical upper limits.

\begin{table}[t]
\centering
\caption{Bayes factors for targeted cluster searches, comparing the CRN+CGW model to the CRN-only model, under eccentric (\texttt{GWecc}) and circular waveform models.}
\label{tab:bf_clusters}
\begin{tabular}{lcc}
\hline
Cluster & eccentric & circular \\
\hline
Hercules & 0.61 & 0.72 \\
Coma     & 1.36 & 0.59 \\
Norma    & 0.90 & 0.49 \\
Fornax   & 0.70 & 0.31 \\
Virgo    & 0.45 & 0.68 \\
\hline
\end{tabular}
\end{table}

\begin{figure}[t]
    \centering
    \includegraphics[width=0.98\linewidth]{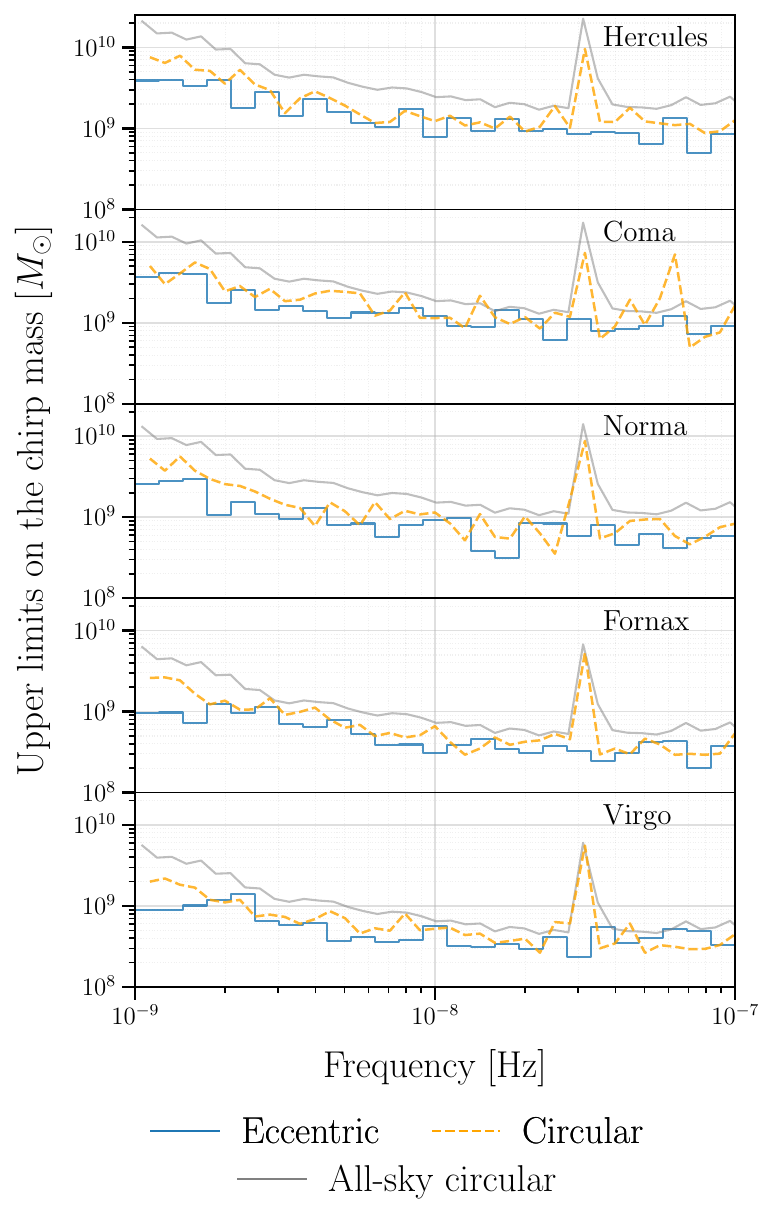}
    \caption{95\% credibility upper limits on the chirp mass as a function of GW frequency for the five targeted galaxy clusters. Results obtained with the eccentric waveform are shown as solid blue curves, while the circular benchmark is indicated by dashed orange curves. Also shown in gray are the upper limits rescaled to the luminosity distance of each galaxy cluster based on the all-sky circular search in Ref. \cite{Zhao2025PPTADR3AllSkyCW}.}
    \label{fig:cluster_mcul}
\end{figure}
In the absence of a detection, we report 95\% credibility upper limits on the chirp mass $\mathcal{M}$ as a function of GW frequency for each cluster (Figure~\ref{fig:cluster_mcul}). As expected from the amplitude--distance scaling $h_0\propto \mathcal{M}^{5/3} f^{2/3}/D_L$, the nearest clusters provide the tightest constraints.
For instance, the constraints for the nearby Virgo cluster ($D_L \approx 16.5$ Mpc) are approximately a factor of four stricter than those for the more distant Hercules cluster ($D_L \approx 160$ Mpc). This aligns with the theoretical expectation where the chirp mass limit scales as $\mathcal{M} \propto D_L^{3/5}$ for a fixed strain sensitivity.
The PPTA DR3 sensitivity peaks at $\sim 20$ nHz. This ``sweet spot" represents an optimal transition window: at lower frequencies, sensitivity is degraded by steep red noise and the absorption of signal power by timing-model fits, while higher frequencies are limited by the white noise floor.
Quantitatively, in the array's most sensitive frequency band, we exclude binary systems with chirp masses $\mathcal{M} \gtrsim 3 \times 10^8 M_{\odot}$ in the Virgo cluster, whereas the corresponding limit for the Hercules cluster is $\mathcal{M} \gtrsim 1 \times 10^9 M_{\odot}$.

These targeted limits are typically a factor of a few more sensitive than those derived from all-sky searches for circular binaries\ucite{Zhao2025PPTADR3AllSkyCW}; shown as gray curve in Figure~\ref{fig:cluster_mcul}.
As a concrete reference point, evaluating at 10~nHz we find that the targeted chirp-mass upper limits are improved by $\sim$2 on average relative to the all-sky circular curve, corresponding to a strain improvement of $\sim 3.2$ (since $h_0 \propto \mathcal{M}^{5/3}$); this is broadly consistent with the level of improvement reported in recent NANOGrav targeted-search studies\ucite{ng_target}.
The limits derived from the eccentric waveform and the circular benchmark are broadly comparable, with differences of $\lesssim 2$ attributable to the eccentric model's richer harmonic structure and associated parameter degeneracies. The generally smoother frequency dependence of the eccentric limits, compared to the circular case, results from the spreading of signal power across multiple harmonics and the marginalization over additional parameters like eccentricity, which dilutes sharp sensitivity features.


\begin{figure*}[t]
    \centering
    \includegraphics[width=0.98\linewidth]{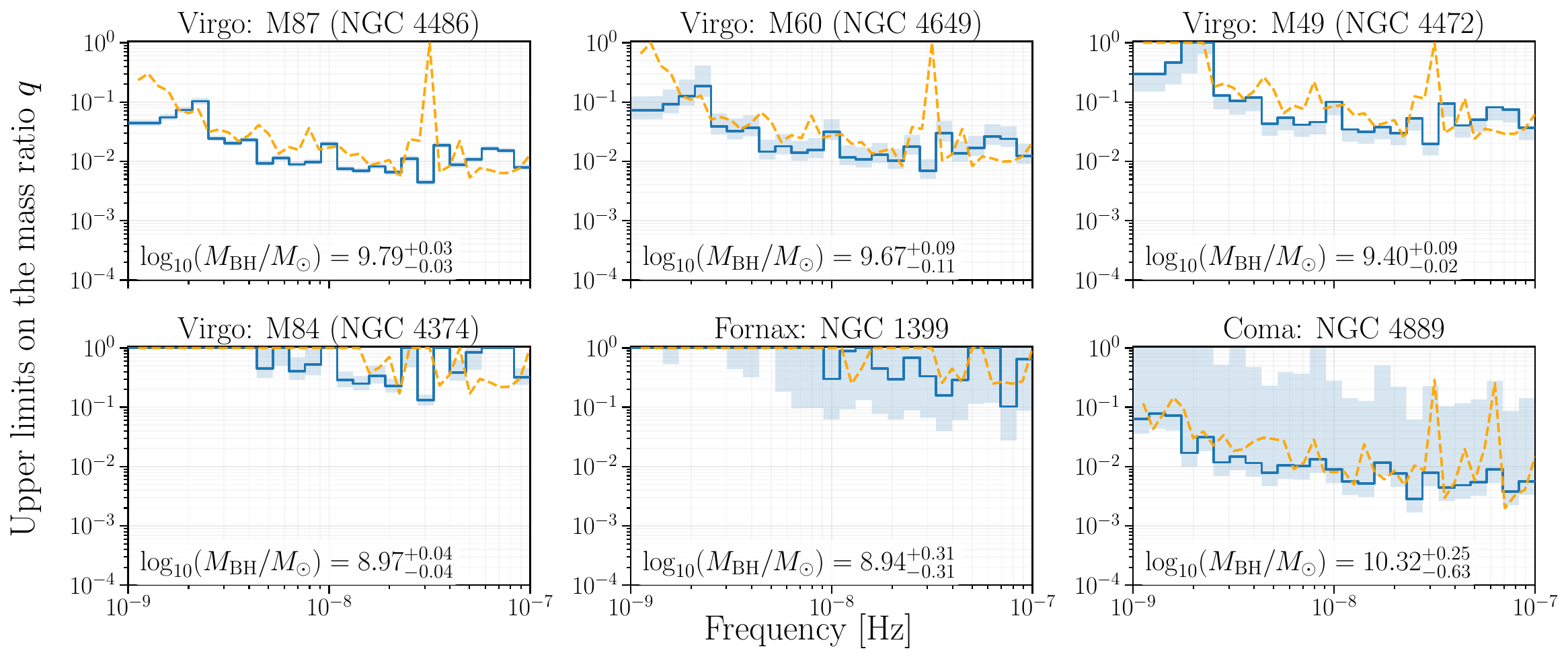}
    \caption{95\% credibility upper limits on the binary mass ratio $q$ as a function of GW frequency for representative central galaxies in the targeted clusters.
    Each panel corresponds to a host galaxy with an adopted central black hole mass $M_{\rm BH}$ (annotated).
    Solid blue curves show the eccentric (\texttt{GWecc}) results and dashed orange curves show the circular benchmark. The shaded bands indicate the 68\% confidence interval in the derived $q$ limits due to the uncertainty in the assumed host black hole mass.}
    \label{fig:q_ul}
\end{figure*}



While PTA data directly constrain the GW amplitude (and thus the chirp mass), these limits can be translated into constraints on the binary mass ratio $q\equiv m_2/m_1\le 1$ for specific galaxies with independent black hole mass estimates $M_{\rm BH}$.
Assuming the total binary mass $M_{\rm tot}$ is comparable to the host SMBH mass, the chirp-mass upper limit $\mathcal{M}^{95\%}$ implies an upper limit on $q$:
\begin{equation}
\eta = \frac{q}{(1+q)^2} = \left(\frac{\mathcal{M}^{95\%}}{M_{\rm tot}}\right)^{5/3}.
\end{equation}
To translate the cluster-wide chirp mass limits to specific mass-ratio constraints for individual member galaxies, we assume that the putative SMBBH is spatially co-located with the host galaxy. Given the large distances involved, we adopt the cluster's central sky position and luminosity distance as a robust proxy for all massive galaxies associated with that cluster.

We present two complementary views of these constraints.
First, Figure \ref{fig:q_ul} shows frequency-dependent mass-ratio limits for representative massive galaxies with dynamical SMBH mass measurements, including M87, M60, M49, and M84 in Virgo, NGC~1399 in Fornax, and NGC~4889 in Coma.
As anticipated, galaxies hosting more massive central black holes yield tighter constraints, as a given chirp-mass limit excludes a broader range of near-equal-mass configurations.

\begin{figure}[t]
    \centering
    \includegraphics[width=\linewidth]{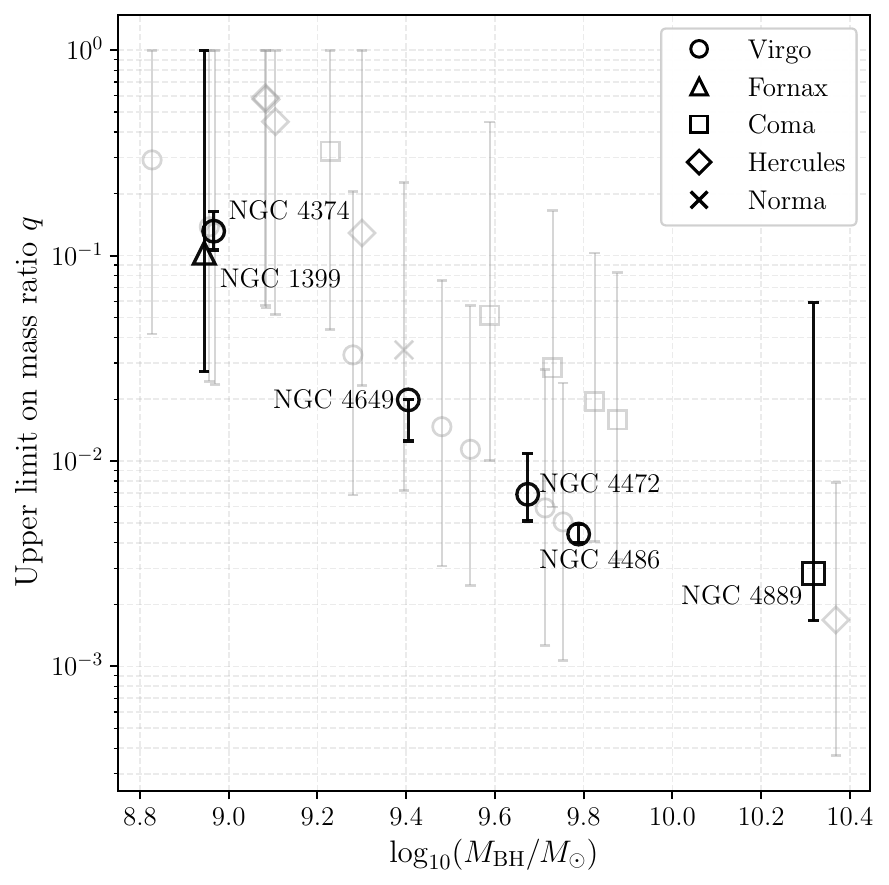}
    \caption{
    Upper limits on the binary mass ratio $q$ at the best frequency for each targeted cluster. For each cluster, we select the frequency bin where the upper limit on the chirp mass is lowest (i.e., the point of greatest sensitivity for the eccentric search in Figure \ref{fig:cluster_mcul}). Black symbols denote galaxies with direct dynamical black hole mass measurements; gray symbols denote galaxies with masses estimated from the $M$--$\sigma$ scaling relation. The vertical error bars show the 68\% confidence interval in the derived $q$ limits, accounting for uncertainties of black hole masses.
    }
    \label{fig:q_ul_galaxies}
\end{figure}

Second, we show in Figure ~\ref{fig:q_ul_galaxies} mass-ratio limits at the most sensitive frequency for each cluster. For every cluster, we identify the frequency bin where our chirp-mass sensitivity is highest (the “best frequency”) and compute the corresponding upper limit for member galaxies (using SMBH masses compiled in Table \ref{tab:galaxy_mbh_cut}). This provides an intuitive, astrophysically interpretable set of constraints, demonstrating that our non-detections already exclude equal-mass binaries over a range of orbital periods for many massive galaxies in these dense environments.

In Figure ~\ref{fig:q_ul_galaxies}, black symbols correspond to galaxies with dynamical black hole mass measurements, while grey symbols use black hole masses inferred from scaling relations. Galaxy membership information and source positions are compiled via \textsc{VizieR}\footnote{\url{https://vizier.cds.unistra.fr/}}, and velocity dispersions $\sigma$ are taken from \textsc{SDSS} SkyServer spectroscopy\ucite{SDSSDR17} where available\footnote{\url{https://skyserver.sdss.org/}}. For the $M$--$\sigma$ derived masses we adopt the following relation\ucite{M_sigma},
\begin{equation}
\log_{10}\!\left(\frac{M_{\rm BH}}{M_\odot}\right)
=
\alpha
+
\beta \log_{10}\!\left(\frac{\sigma}{200~{\rm km~s^{-1}}}\right),
\end{equation}
with $\alpha=8.32$ and $\beta=5.64$.
We caution that the $M-\sigma$ relation used here carry intrinsic scatter of typically $0.3-0.5$ dex. For the grey symbols in Figure~\ref{fig:q_ul_galaxies}, the vertical error bars are obtained by propagating a fiducial 0.4 dex uncertainty in $\log_{10} M_{\rm BH}$ associated with the scaling relation, while for the black symbols we use the published uncertainties of the dynamical black hole mass measurements. These systematic uncertainties in the assumed primary mass $M_{\mathrm{BH}}$ propagate to the derived mass ratio limits roughly as $q \propto M_{\mathrm{BH}}^{-5/3}$; consequently, a factor of two uncertainty in the primary mass can shift the upper limit on $q$ by a factor of a few. Constraints for these indirect targets should therefore be interpreted as order-of-magnitude estimates.

Taking M87 as an example, we discuss the astrophysical implications of our results in more detail.
Our targeted PTA limits on the binary mass ratio for M87 ($q \lesssim 0.01$ at periods of $\sim 3-10$ years) offer a complementary probe to the horizon-scale constraints derived from the Event Horizon Telescope (EHT). While PTAs are sensitive to the dynamics of SMBBHs in the parsec-scale inspiral regime ($a \sim 0.01-0.1$ pc), mm-VLBI imaging probes the immediate vicinity of the primary black hole at milliparsec scales ($a \lesssim 0.005$ pc). For instance, Ref. \cite{safarzadeh2019M87} demonstrated that the persistence of the annular shadow structure observed by the EHT effectively excludes a massive companion ($q \gtrsim 0.1$) at separations comparable to the shadow diameter, as such a perturber would tidally disrupt the accretion flow and distort the photon ring. Furthermore, high-precision centroid astrometry from the EHT can constrain the `wobble' of the primary black hole induced by a companion, ruling out significant regions of the $q-a$ parameter space that are inaccessible to current GW searches. Consequently, the combination of our nHz GW upper limits and mm-VLBI imaging data provides a multi-scale exclusion of major merger remnants in M87, effectively disfavoring equal-mass binaries from the parsec-scale inspiral down to the near-horizon regime.

\medskip

\textbf{Conclusions.}

We have performed targeted Bayesian searches for CGWs from eccentric binaries toward the blazar OJ 287 and the centers of five nearby galaxy clusters (Virgo, Fornax, Norma, Hercules, and Coma) using the PPTA DR3 dataset. Our analysis explicitly accounts for eccentric inspirals and the pulsar term, ensuring robustness against waveform systematics that could affect standard quasi-circular searches.

For the blazar OJ 287, we find no statistically significant evidence for a CGW signal, with the Bayes factor for the eccentric model consistent with the noise-only hypothesis ($\mathcal{B} \approx 0.82$). Accordingly, we place a 95\% credibility upper limit on the total binary mass of $M_{\rm tot} < 5.25 \times 10^{10} M_{\odot}$. Our analysis indicates that for this target, current PTA sensitivity does not significantly update the validity-restricted prior except in the highest mass regime, and our strain upper limits remain above the predictions of current EM binary models.
However, establishing this direct model-independent GW limit is a critical milestone. As PTA sensitivities improve over the coming years, these direct measurements will eventually become stringent enough to distinguish between competing EM interaction and flare-timing models proposed for OJ 287.

In the searches toward the five galaxy clusters, the Bayes factors generally favor the null hypothesis, with the Coma cluster showing a mild but statistically insignificant preference for a signal ($\mathcal{B} \approx 1.36$). Treating these as non-detections, we derived frequency-dependent upper limits on the chirp mass. By combining these limits with independent black hole mass estimates, we placed novel constraints on the binary mass ratio, $q$, for potential host galaxies within these clusters. Notably, for the central galaxies M87 (Virgo) and NGC~4889 (Coma), our results exclude the presence of equal-mass binaries ($q \approx 1$) across a broad range of nanohertz frequencies, thereby beginning to constrain major-merger SMBBHs in these massive nearby hosts.

These results highlight the growing power of PTAs to probe SMBBH populations in the local Universe. In particular, the ability to translate PTA non-detections into constraints on binary mass ratio provides an astrophysically interpretable way to test whether nearby massive galaxies could still host near-equal-mass SMBBHs. Furthermore, our constraints on $q$ in the parsec-scale inspiral regime also offer a complementary ``outer'' boundary to the ``inner'' horizon-scale constraints derived from millimeter-VLBI imaging (e.g., by the EHT), illustrating how PTA and EM observations can jointly probe SMBBHs across very different physical scales. Future PTA data releases, featuring extended temporal baselines and improved noise characterization, will be essential to further tighten these astrophysical limits and enable more stringent tests of eccentric SMBBH populations in dense environments such as galaxy clusters.

\medskip

\textit{Acknowledgements.} This work is supported by the National Key Research and Development Program of China (No. 2023YFC2206704), the Fundamental Research Funds for the Central Universities, and the Supplemental Funds for Major Scientific Research Projects of Beijing Normal University (Zhuhai) under Project ZHPT2025001. 
Parts of this work was completed with support from the Australian Research Council (ARC) Centre of Excellence CE230100016.
We thank Abhimanyu Susobhanan and Achamveedu Gopakumar for useful discussions. The Parkes radio telescope (Murriyang) is part of the Australia Telescope National Facility which is funded by the Australian Government for operation as a National Facility managed by CSIRO. We acknowledge the Wiradjuri People as the traditional owners of the Observatory site.
This paper includes archived data obtained through the Parkes Pulsar Data archive on the CSIRO Data Access Portal 
(http://data.csiro.au).
This work was performed on the OzSTAR national facility at Swinburne University of Technology. The OzSTAR program receives funding in part from the Astronomy National Collaborative Research Infrastructure Strategy (NCRIS) allocation provided by the Australian Government, and from the Victorian Higher Education State Investment Fund (VHESIF) provided by the Victorian Government.


\medskip


\onecolumn

\begin{center}
{\large\bfseries Appendix A: Catalog of galactic central black hole mass estimates}
\end{center}

Table \ref{tab:galaxy_mbh_cut} compiles the catalog of supermassive black hole masses assumed for member galaxies across the five targeted clusters: Virgo, Fornax, Coma, Hercules, and Norma. These mass estimates provide the necessary baseline for converting the PTA-derived chirp mass limits into the astrophysical mass-ratio constraints presented in Figure \ref{fig:q_ul_galaxies}. The table explicitly distinguishes the source of these estimates, identifying galaxies with robust direct dynamical measurements—such as M87 and NGC 4889—with an asterisk, while the remaining values are derived indirectly using scaling relations.

\begingroup
\small
\setlength{\tabcolsep}{3pt}

\begin{longtable}{l @{\hspace{0.0em}} c @{\hspace{0.2cm}}
                  l @{\hspace{0.0em}} c}
\caption{Black hole masses used for the informative mass-ratio $q$ limits for massive galaxies in targeted clusters (Figure ~\ref{fig:q_ul_galaxies}). Galaxies with direct dynamical black hole mass measurements are marked with a superscript ``$^\ast$''; all others are derived using the $M$--$\sigma$ scaling relation. For scaling-relation-based masses, we adopt a fiducial uncertainty of $\pm 0.40$ dex in $\log_{10}(M_{\rm BH}/M_\odot)$ when propagating the uncertainty to the derived $q$ limits.}
\label{tab:galaxy_mbh_cut}\\
\hline
Galaxy ID & $\log_{10}(M_{\rm BH}/M_\odot)$ &
Galaxy ID & $\log_{10}(M_{\rm BH}/M_\odot)$ \\
\hline
\endfirsthead

\hline
Galaxy ID & $\log_{10}(M_{\rm BH}/M_\odot)$ &
Galaxy ID & $\log_{10}(M_{\rm BH}/M_\odot)$ \\
\hline
\endhead

\hline
\multicolumn{4}{r}{\footnotesize Continued on next page.}\\
\hline
\endfoot

\hline
\endlastfoot

\multicolumn{4}{l}{\textbf{Virgo}} \\
M87 (NGC 4486)$^\ast$ & $9.789^{+0.026}_{-0.027}$ & VCC 1189 & $9.754 \pm 0.40$ \\
SDSSspec-55 & $9.713 \pm 0.40$ & M60 (NGC 4649)$^\ast$ & $9.674^{+0.086}_{-0.109}$ \\
SDSSspec-307 & $9.545 \pm 0.40$ & VCC 1059 & $9.480 \pm 0.40$ \\
M49 (NGC 4472)$^\ast$ & $9.405^{+0.089}_{-0.017}$ & VCC 350 & $9.280 \pm 0.40$ \\
VCC 1224 & $8.969 \pm 0.40$ & M84 (NGC 4374)$^\ast$ & $8.966^{+0.044}_{-0.043}$ \\
VCC 4 & $8.956 \pm 0.40$ & SDSSspec-243 & $8.827 \pm 0.40$ \\
\hline

\multicolumn{4}{l}{\textbf{Fornax}} \\
NGC 1399$^\ast$ & $8.945^{+0.306}_{-0.306}$ &  &  \\
\hline

\multicolumn{4}{l}{\textbf{Coma}} \\
NGC 4889$^\ast$ & $10.318^{+0.245}_{-0.628}$ & JT25-2021 & $9.876 \pm 0.40$ \\
JT25-1935 & $9.825 \pm 0.40$ & JT25-1933 & $9.730 \pm 0.40$ \\
JT25-2038 & $9.588 \pm 0.40$ & JT25-2040 & $9.229 \pm 0.40$ \\
\hline

\multicolumn{4}{l}{\textbf{Hercules}} \\
SDSSspec-37 & $10.368 \pm 0.40$ & SDSSspec-17 & $9.300 \pm 0.40$ \\
SDSSspec-171 & $9.105 \pm 0.40$ & SDSSspec-151 & $9.084 \pm 0.40$ \\
SDSSspec-134 & $9.083 \pm 0.40$ &  &  \\
\hline

\multicolumn{4}{l}{\textbf{Norma}} \\
2MASX J16505460-6148452 & $9.395 \pm 0.40$ &  &  \\
\hline

\end{longtable}
\endgroup

\end{document}